\documentclass[pre, aps, preprint, onecolumn, superscriptaddress, nofootinbib, nobibnotes, showpacs, 11pt]{revtex4-1}

\usepackage{natbib}
\usepackage{slashed}
\usepackage{graphicx}
\usepackage{subfigure}
\usepackage[usenames, dvipsnames]{color}
\usepackage{graphics}
\usepackage{hyperref}
\usepackage{bm}
\usepackage{comment}
\usepackage{amsmath}
\usepackage{color}\label{key}
\usepackage{amsfonts}
\hypersetup{backref,
colorlinks=true,
linkcolor=blue,
linktoc=page,
citecolor=blue,
urlcolor=blue}

\everymath{\displaystyle}

\begin{document}

\title{Plasmon dispersion and Landau damping in the nonlinear quantum regime}

\author{F. Haas}

\email{fernando.haas@ufrgs.br}

\affiliation{Physics Institute, Federal University of Rio Grande do Sul, Av. Bento Gon\c calves 9500, 91501-970 Porto Alegre RS, Brazil.}

\author{J. T. Mendon\c{c}a}

\email{titomend@tecnico.ulisboa.pt}

\affiliation{GoLP / IPFN, Instituto Superior T\'{e}cnico, Universidade de Lisboa, Av. Rovisco Pais 1, 1049-001 Lisboa, Portugal.}

\author{H. Ter\c cas}

\email{hugo.tercas@tecnico.ulisboa.pt}

\affiliation{GoLP / IPFN, Instituto Superior T\'{e}cnico, Universidade de Lisboa, Av. Rovisco Pais 1, 1049-001 Lisboa, Portugal.}

\begin{abstract}

We study the dispersion properties of electron plasma waves, or plasmons, which can be excited in quantum plasmas in the nonlinear regime. In order to describe nonlinear electron response to finite amplitude plasmons, we apply the Volkov approach to non-relativistic electrons. For that purpose, we use the Schr\"odinger equation and describe the electron population of a quantum plasma as a mixture of quantum states. Within the kinetic framework that we are able to derive from the Volkov solutions, we discuss the role of the wave amplitude on the nonlinear plasma response. Finally, we focus on the quantum properties of nonlinear Landau damping and study the contributions of multi-plasmon absorption and emission processes. \end{abstract}

\maketitle

\section{Introduction}

Landau damping of electron plasma waves was introduced in 1946, in the frame of classical plasma physics, and plays a central role in plasma theory \cite{landau1946, Dawson1961, Nicholson1983, Weiland1981, Ryutov1999}. Although quite well understood in its linear formulation, the nonlinear regime of Landau damping remains a problem of major mathematical and physical importance \cite{Villani2002, Zhou2001, Shaokang2021}, not only in plasmas but also in self-gravitating systems \cite{Wiechen1988, Moretti2020}. Its quantum version was introduced in the early 60's of the last century \cite{Klimontovich1960, Pines1962}, and seems to play a more modest role in quantum plasma theory \cite{Manfredi2001,Haas2011}. Despite some progress in more recent years \cite{Daligault2014, Brodin2015, Chatterjee2016}, the nonlinear effects associated with quantum Landau damping, such as those associated with particle trapping, wave satellites and field harmonics, are still not well understood. While the basic properties of quantum Landau damping can be found in books \cite{Haas2011} and review articles \cite{Misra2022}, a variety of effects associated with the quantum nature of the plasmon excitations remain to be tackled. For instance, it has been recently reported numerically that quantum Landau damping consists of a {\it multi-plasmon damping}, and it can be understood as a balance between emission and absorption of plasmons by the plasma electrons \cite{Brodin2017}.

In this paper, we propose an innovative approach to quantum Landau damping in the nonlinear regime. Our method is based on the used of Volkov (also known as Wolkow \cite{note}) solutions to describe the single particle (test electron) states in the presence of an electrostatic wave. A remarkable feature of our approach is the fact of being, by construction, nonperturbative in the wave amplitude. Volkov solutions to the Dirac equation were originally obtained for relativistic quantum particles in vacuum \cite{Wolkow1935, Itzykson2006}. In recent years, they have been extended to plasmas, where approximate solutions can also be found \cite{Mendonca2011, Shalom2013, Varro2013}. These solutions can be used to describe multi-photon and multi-plasmon effects associated with inverse bremsstrahlung \cite{Mendonca2013} and Compton scattering \cite{Brown1964, Mendonca2023b}. In the following, we consider the non-relativistic quantum plasma case, where the Dirac equation is replaced by the Schr\"odinger equation. \par
Previous studies of Volkov solutions of the Schr\"odinger equation have focused on single particle states in vacuum \cite{Reiss1994, Bauer2002, Li2020}. Here, instead, we consider an electron plasma wave with a finite amplitude. We use the single-particle Volkov solutions in the presence of this wave and establish the electron plasma population as a mixture of quantum states. This allows us to describe the nonlinear plasma response to the electron plasma wave. Such a description allows us to derive a new dispersion relation where multi-plasmon effects are included. We then focus on the quantum properties of nonlinear Landau damping, and study the contributions of multi-plasmon absorption and emission processes. Our analytical results confirm and extend previous work, mainly based on numerical solutions \cite{Brodin2015, Brodin2017}, and propose a new conceptual approach to nonlinear plasma. Our results may certainly contribute to enlarge the discussion around wave-particle effects in nonlinear waves in quantum systems, namely in the context of graphene devices \cite{Hwang2007, Badikova2017, Cosme2020, Cosme2021}. \par
This paper is organized as follows. In Section II, we introduce exact single-electron solutions driven by the plasma wave with the help of the Volkov formalism. Then, in Section III, we obtain a nonlinear dispersion relation for a generic equilibrium based on the quantum (Wigner) kinetic equation. Finally, in Sec. IV some conclusions are stated and applications in topical physical problems are motivated.

\section{Volkov solutions for non-relativistic electrons}

We start from Schr\"odinger's equation describing single-electron states in the presence of an electrostatic potential, as
\begin{equation}
i \hbar \frac{\partial}{\partial t} \psi = \left[ - \frac{\hbar^2}{2 m} \nabla^2 + e V \right] \psi \, ,
\label{2.1} 
\end{equation} 
where $\psi \equiv \psi ({\bf r}, t)$ is the electron wavefunction and $V \equiv  V ({\bf r}, t)$ is the electrostatic potential associated with an electron plasma wave with frequency $\omega$ and wavevector ${\bf k}$ that propagates in the plasma. We now look for quantities varying in terms of the {\it proper time} $\tau = t - {\bf k} \cdot {\bf r} /\omega$, such that $V ({\bf r}, t) = V_0 f (\tau)$, where $V_0$ is the wave amplitude, and the function $f (\tau)$ describes the wave shape to be specified. Following the Volkov procedure, we assume a solution of the form 
\begin{equation}
\psi ({\bf r}, t) = \Phi (\tau) \exp (i \theta) \, , \quad \theta =  {\bf k}_e \cdot {\bf r} - \omega_e t  .
\label{2.2} 
\end{equation}
Here, ${\bf p}_e = \hbar {\bf k}_e$ is the electron momentum and $\hbar \omega_e = p_e^2 / 2 m$ the corresponding kinetic energy. This allows us to use
\begin{equation}
\frac{\partial \psi}{\partial t} = \left( \Phi' - i \omega_e \Phi \right) \exp ( i \theta) \, ,
\label{2.2b} 
\end{equation}
with $\Phi' \equiv \frac{\partial \Phi}{\partial t} = \frac{\partial \Phi}{\partial \tau}$. Similarly, for the space derivatives, we have
\begin{equation}
\nabla \psi = \left( i {\bf k}_e \Phi - \frac{{\bf k}}{\omega} \Phi' \right) \exp ( i \theta) \, ,
\label{2.3} 
\end{equation}
and 
\begin{equation}
\nabla^2 \psi = \left( \frac{k^2}{\omega^2} \Phi'' - 2 i \frac{({\bf k}_e \cdot {\bf k})}{\omega} \Phi' - k_e^2 \Phi \right) \exp (i \theta) \, .
\label{2.3b} 
\end{equation}
Replacing this in Eq. (\ref{2.1}), we get an evolution equation for $\Phi$, of the form
\begin{equation}
\alpha^2 \Phi'' + i g \Phi' - G (\tau) \Phi = 0 \, , 
\label{2.4} 
\end{equation}
with the following quantities
\begin{equation}
\alpha^2 = \frac{\hbar^2 k^2}{2 m \omega^2} \, , \quad g = \hbar \left(1 - \frac{{\bf p}_e \cdot {\bf k}}{m \omega} \right) \, , \quad
G (\tau ) = e V_0 f (\tau) \, .
\label{2.4b} 
\end{equation}
These auxiliary quantities have a very clear physical meaning: $\alpha^2$ represents the kinetic energy of a particle with the wave momentum $\hbar {\bf k}$ (the factor $\omega^2$ appear as to weight the second time derivative); the factor $g$ represents the deviation of the particle velocity with respect to the wave phase velocity; and the function $G (\tau)$ is simply the potential energy. Eq. (\ref{2.4}) can now be converted into a first order equation using
\begin{equation}
\Phi ( \tau ) = \Phi_0 \exp \left[ - i \int_0^\tau \Omega ( \tau ' ) d \tau ' \right] ,
\label{2.5} 
\end{equation} 
where $\Phi_0\equiv\Phi(\tau=0)$ is a constant. We get the Riccati equation
\begin{equation}
- \alpha^2 \left(  i \Omega' + \Omega^2 \right) + g \Omega - G (\tau) = 0.
\label{2.5b} 
\end{equation}
We now assume a sinusoidal electrostatic wave in the medium, such that
$G (\tau )= e V_0 \exp (- i \omega \tau)$. To the lowest order in the wave amplitude $V_0$, we can neglect the nonlinear term, we can derive the following particular solution for the eikonal phase
\begin{equation}
\Omega(\tau) = \sum_{n=1}^\infty \Omega_{n}e^{- i n \omega \tau} , 
\label{2.6} 
\end{equation}
with the first coefficients being given as
\begin{equation}
\Omega_1 = \frac{e V_0}{g - \omega \alpha^2}, \quad 
\Omega_2 = \frac{\alpha^2}{g-2\omega\alpha^2} \left(\frac{e V_0}{g - \omega \alpha^2} \right)^2, \quad 
\Omega_3 = \frac{2\alpha^4}{(g-2\omega\alpha^2)(g-3\omega\alpha^2)} \left(\frac{e V_0}{g - \omega \alpha^2} \right)^4,  ~ \ldots .
\end{equation}
The formal particular solution in Eq. \eqref{2.6} is not very useful for the collective treatment we will perform below. As such, in order to cast the nonlinear effects at first order the in the wave amplitude, we neglect terms of the order $\mathcal{O}(V_0^2)$. In order to avoid unbounded (exponentially growing) solutions, we take the real part of the phase in Eq. \eqref{2.6}. Consequently, the formal solution to Eq. \eqref{2.5} can be given in terms of the Jacobi-Anger expansion, yielding the following bounded functions
\begin{equation} 
\Phi (\tau ) \simeq \Phi_0 \sum_\ell i^\ell J_\ell (\xi ) e ^{- i \ell \omega \tau},  
\label{2.6b} 
\end{equation}
where $\xi=\Omega_1/\omega$ plays the role of a dimensionless electrostatic energy. This shows that the single electron states include multi-plasmon transitions (emission and absorption), that are weighted by the wave amplitude $V_0$, with probability $\left| J_\ell (\xi ) \right|^2$. Such effects will be explored below. This is justified as we are interested in investigating stable solutions only. We notice, nevertheless, that an alternative and more generic treatment of the problem would require the solutions of the Mathieu 
\begin{equation}
\alpha^2\varphi ''+\left( \frac{g^2}{4\alpha^2}-G(\tau)\right)\varphi=0,
\end{equation}
obtained from Eq. \eqref{2.4} upon the transformation $\varphi(\tau) = \exp[ig \tau/(2\alpha^2)]\phi(\tau)$. However, this would lead to formal difficulties preventing our analysis, since it does explicitly show the multi-plasmon characteristic of the nonlinear solution of Eq. \eqref{2.6b}, as it will become apparent below. In the remainder of this work, we are interested in recasting the effect of the finite amplitude in the dispersion of the waves, and not in investigating the solutions of single electrons.  
\begin{figure}
\includegraphics[width=\columnwidth]{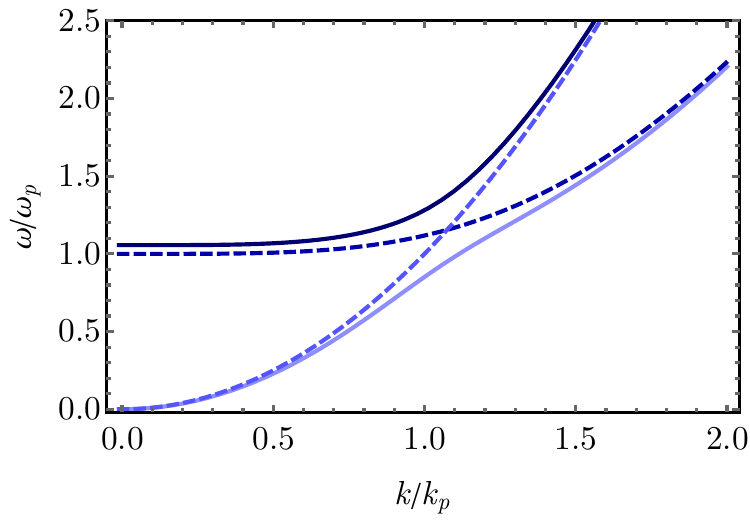}
\caption{(color online) {\bf Dispersive properties of electron plasma waves with finite amplitude}. The dark dashed line corresponds to the linear dispersion relation obtained for $V_0\to 0$, $\omega=\sqrt{\omega_p^2+\nu_k^2}$, while the light dashed line is the single particle dispersion $\omega=\hbar k^2/m$ (see text). The upper (dark solid line) and lower (light solid line) are the nonlinear plasma modes obtained for a wave of amplitude $V_0=0.2 \hbar \omega_p/e$. The two nonlinear modes repel each other at the crossing point of the linear modes, $k_*=(4/3)^{1/4}k_p$, with $k_p=\sqrt{\omega_p/\hbar^2 m}$ defining the scale of the oscillations. }
\label{fig_real}
\end{figure}

\section{Nonlinear plasmons in a quantum plasma}

According to Eqs. (\ref{2.2}) and (\ref{2.6b}), we can characterize a given electron state with velocity ${\bf v}={\bf p}_e/m$, in the presence of a finite amplitude plasma wave, by a wavefunction of the form
 \begin{equation}
\psi_{\bf v} ({\bf r}, t) = \Phi_0  \exp (i \theta_{\bf v}) \sum_\ell i^\ell J_\ell (\xi ) \exp (- i \ell \omega \tau)  \, .
\label{3.1} \end{equation}
and describe the quantum fluid as a mixture of states \cite{Haas2011} with quasi-probability
\begin{equation}
W ({\bf r}, {\bf v}, t ) = \int d{\bf v'} W_0 ({\bf r}, {\bf v'}-{\bf v}) W_{\bf v'} ({\bf r}, t)  ,
\label{3.1b} 
\end{equation}
where $\theta_{\mathbf{v}}=m{\bf v}\cdot{\bf r}/\hbar-\omega_e t$ is the test electron phase (the subscript ${\bf v}$ is simply introduced to render the dependence on the electron velocity ${\bf v}$ more explicit), $W_0 ({\bf r}, {\bf v})$ is the equilibrium distribution at a given electron temperature $T_e$ (typically, the Fermi-Dirac distribution), and the functions 
$W_{\bf v} ({\bf r}, t)$ are Wigner functions defined by
\begin{equation}
W_{\bf v} ({\bf r}, t) = \int \psi_{\bf v} ({\bf r} - {\bf s}/2, t) \psi_{\bf v}^* ({\bf r} + {\bf s}/2, t) e^ { i m{\bf v} \cdot {\bf s}/\hbar} \, d {\bf s} .
\label{3.2} 
\end{equation}
Let us now study the plasma dispersion using this nonlinear function of the wave amplitude as the undisturbed distribution. 
For this purpose, we use the wave-kinetic equation for a quantum plasma \cite{Haas2011}
\begin{align}
i \hbar \left( \frac{\partial}{\partial t} + {\bf v} \cdot \bm\nabla \right) W = e \int  d {\bf q}  V ({\bf q}) \left[ W^- - W^+ \right] e^{i {\bf q} \cdot {\bf r}}\, ,
\label{3.2b} 
\end{align}
where $V({\bf q})$ is the Fourier transform of the electrostatic potential created by the plasma, $W^\pm = W ({\bf {r}}, {\bf v} \pm {\bf v_q},t )$, and ${\bf v_q} = \hbar {\bf q} / 2 m$. Here, we have
\begin{align}
\int d {\bf q} V ( {\bf q}) W ({\bf r} ,{\bf v} \pm {\bf v_q},t) e^{i {\bf q} \cdot {\bf r}}  &= \sum_\ell \int d {\bf q}    V ( {\bf q}) W_0 ({\bf r}, {\bf v} \pm {\bf v_q} ) e^{i {\bf q} \cdot {\bf r}}\int d {\bf s}  \left| J_\ell (\xi) \right|^2
e^{i  m(\ell+1) {\bf v} \cdot ({\bf r} \pm {\bf s})/\hbar} \\
& = \sum_\ell \int d {\bf q}    V ( {\bf q}) W_0 \left[{\bf r}, {\bf v} \pm (\ell +1){\bf v_q} \right]  e^{i {\bf q} \cdot {\bf r}} \left| J_\ell (\xi) \right|^2.
\label{3.2c} 
\end{align}
Equation \eqref{3.2b} must be consistent with the Poisson equation
\begin{equation}
\nabla^2 V({\bf r},t)=\frac{e}{\epsilon_0}\left(\vert \psi ({\bf r},t) \vert^2-n_0(\bf r)\right) =\frac{e}{\epsilon_0} \int \left[W({\bf r},{\bf v},t)-W_0({\bf r}, {\bf v}) \right] d{\bf v},
\end{equation}
with $n_0(\bf r)$ being the background ionic density forcing the plasma quasi-neutrality. In order to obtain the kinetic dispersion relation, we perturb the Wigner-Poisson system as $W=W_0+\tilde W$ and $V=\tilde V$, evolving as $(\tilde W, \tilde V)\sim \exp (i {\bf k} \cdot {\bf r} - i \omega t)$. Using the usual properties of the delta function in Eq. \eqref{3.2c}, and replacing the dummy variables for the wavevector $\bf{q}\to \bf{k}$, we obtain, for the case of a homogenous equilibrium $W_0({\bf r},{\bf v})=n_0G_0({\bf v})$,
\begin{equation}
- i \left( \omega - {\bf k} \cdot {\bf v} \right) \tilde W = n_0 \sum_\ell G_0 \left[{\bf v} - (\ell+ 1) {\bf v_k}\right]  \left| J_\ell (\xi) \right|^2 \tilde V \, ,
\label{3.3} 
\end{equation}
and 
\begin{equation}
- k^2 \tilde V = \frac{e}{\epsilon_0} \int \tilde W d {\bf v} .
\label{3.3b} 
\end{equation}
Finally, combining the latter, we get the nonlinear dispersion relation,
\begin{equation}
1  - \frac{m \omega_p^2}{ \hbar k^2} \sum_{\ell=0}^\infty  \int \left| J_\ell (\xi) \right|^2 \frac{G_0 \left[{\bf v} - (\ell +1) {\bf v_k}\right] - G_0 \left[{\bf v} + (\ell+1) {\bf v_k}\right]}{(\omega - {\bf k} \cdot {\bf v})} d {\bf v} = 0  \, .
\label{3.4} 
\end{equation}
Integrating over the perpendicular velocity ${\bf v}_\perp$, and defining the parallel Wigner function
\begin{equation}
G_0 ( u ) =  \int G_0 ( u, {\bf v}_\perp)  d {\bf v}_\perp \, ,
\label{3.4b} \end{equation}
where $u$ is the particle velocity component along the direction of wave propagation, we get
\begin{equation}
1 - \omega_p^2 \sum_\ell (\ell + 1) \int  \left\vert J_\ell (\xi) \right \vert^2\frac{ G_0 ( u ) d u}{(\omega - k u)^2 - (\ell+1)^2 \nu_k^2} = 0, 
\label{3.5} 
\end{equation}
where $\nu_k=\hbar k^2/2m$ is the recoil frequency. This is a generalization of the usual quantum dispersion, when the Volkov solutions are taken into account. In the limit of infinitesimal wave amplitudes, $\xi\to 0$, we have
\begin{equation}
J_0 (\xi ) \rightarrow 1 \, , \quad J_{\ell \neq 0} ( \xi ) \rightarrow 0 \, , 
\label{3.5b} \end{equation}
and we are reduced to the usual linear dispersion relation of quantum plasmas, as obtained within the random phase approximation (RPA) formalism \cite{Moldabekov2018, Haas2011}
\begin{equation}
1 - \omega_p^2 \int \frac{ G_0 ( u ) d u}{(\omega - k u)^2 -\nu_k^2} = 0. 
\label{3.5bb} 
\end{equation}
In the cold plasma approximation, where we can use $G_0 ( u ) = n_0 \delta ( u )$, the leading order correction to the plasmon dispersion expression is implicitly given by
\begin{equation}
\omega^2 =  \omega_p^2 \sum_\ell (\ell+1) J_\ell^2\left(\xi_0\right)\frac{\omega^2-\nu_k^2}{\omega^2-(\ell+1)^2\nu_k^2}+\nu_k^2,
\label{3.6} 
\end{equation}
where $\xi_0\equiv \xi\vert_{{\bf p}_e=0}=e V_0/\hbar\omega_p$. In the linear regime, $\xi_0\to 0$, we simply get $\omega\simeq \sqrt{\omega_p^2+\nu_k^2}$ \cite{Tsintsadze2009, Haas2011}, and at order $\mathcal{O}(\xi_0^2)$, we get two modes, $\omega\simeq\omega_{\pm}$, which, in the long wavelength limit, read 
\begin{equation}
\omega_+ \simeq \omega_p\left(1+\frac{\vert \xi_0\vert ^2}{2}\right)^{1/2}, \quad \omega_-\simeq \nu_k\left(1- \frac{3}{8}\vert \xi_0\vert^2\right).
\label{eq_app}
\end{equation}
This results expresses a remarkable feature, as depicted in Fig. \ref{fig_real}, which is intrinsic to the nonlinear nature of the oscillations in the presence of a finite amplitude wave: the hybridization between the collective (gapped, $\omega\sim\omega_p$) mode and the single-particle (gapless, $\omega\sim k^2$) mode. At the crossing point, $k_*=(4/3)^{1/4}k_p$, with $k_p=\sqrt{\omega_p/\hbar^2 m}$ defining the spatial scale of the oscillations in a quantum plasma, the two modes repeal by the amount of $2\Omega$, where $\Omega=\sqrt{3}\omega_p\vert \xi_0 \vert/4$ plays the role of a {\it Rabi frequency}, in similarity to what happens in quantum optics \cite{Merlin2021}. Indeed, such avoided crossing has been recently identified in magnetized plasmas interacting with axions \cite{Tercas2018}. We attribute this effect to intricate wave-particle interaction that takes place in the nonlinear regime: while part of the electrons participate in the collective mode, oscillating alongside with the wave at frequency $\omega=\omega_+ \simeq \omega_p$ (corrected by the Volkov mode $\ell=0$), others remain trapped inside the wave, featuring essentially single particle motion $\omega=\omega_-\simeq2\nu_k$. A factor of two correcting the value of the frequency, however, appears here as this oscillation involves, at leading order, the exchange of two plasmons, $\ell=\pm 1$. Of course, multiple plasmon exchange is also expected, but such effect is less important for weakly nonlinear waves, for which the $\ell=\pm 1$ mixture is enough to explain the physical picture of the wave-particle interaction. The interplay of multiple plasmons in wave-particle problem has been observed in numerical simulations \cite{Brodin2017}. We also observe that the frequency shift to the linear plasma frequency $\omega_p$ in Eq. \eqref{eq_app} proportional to $| \xi_0 | \propto V_0$. This strongly differs from the classical result, associated to electron trapping at late stages of the wave-particle interaction, in which trapping is proportional to $\sqrt{V_0}$, and not to $V_0$ as we obtain here. \par
\begin{figure}
\includegraphics[width=\columnwidth]{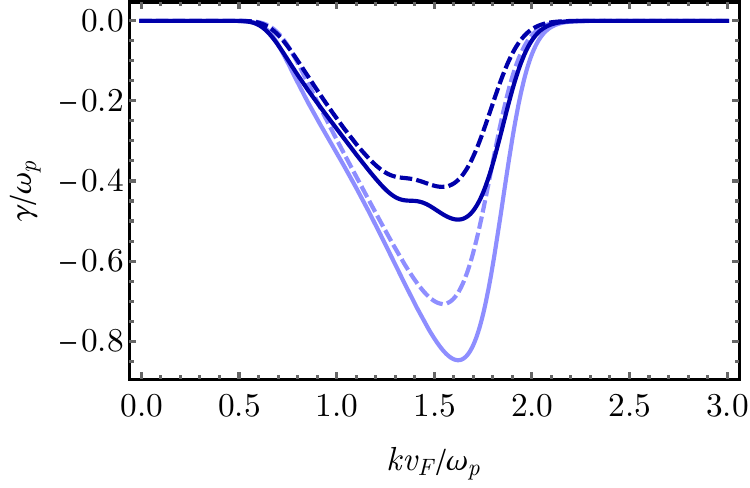}
\caption{(color online) {\bf Plasmon Landau damping in the nonlinear quantum regime.} The lighter blue lines correspond to the linear case $V_0= 0$, while the darker lines are obtained for $V_0=0.2 \hbar\omega_p/e$. In both cases, dashed and solid lines correspond to $E_F=0.75\hbar \omega_p$  and $E_F=1.10\hbar \omega_p$, respectively, showing the suppression of damping due to quantum effects as well. For numerical illustration, we have chosen a degeneracy parameter of $E_F/T_e=5$, and defined $v_F=\sqrt{E_F/m}$. }
\label{fig_imag}
\end{figure}

A more interesting situation concerns the isentropic decay of plasmons in thermal plasmas, $T_e\neq 0$, known as {\it Landau damping}. Coming back to Eqs. \eqref{3.4} and \eqref{3.4b}, we make use of the Landau prescription, in which the integral in $u$ is performed by forcing the contour $\mathcal{L}$ to pass {\it below} the pole $u=\omega/k$ \cite{Manfredi2001, Zhu2009}, and assuming that the frequency acquires a small imaginary part, $\omega\to\omega+i\gamma$, such that
\begin{align}
\epsilon(k,\omega)&=1 - \frac{m\omega_p^2}{\hbar k^2} \sum_\ell (\ell + 1) \int_\mathcal{L}   \left\vert  J_\ell (\xi) \right \vert^2 \frac{G_0 \left[u - (\ell +1) v_k\right] - G_0 \left[u + (\ell+1)  v_k\right]}{\omega +i\gamma  - k  u} d u\\
&\simeq 1 - \frac{m\omega_p^2}{\hbar k^2} \sum_\ell (\ell + 1) {\wp} \int   \left\vert  J_\ell (\xi) \right \vert^2 \frac{G_0 \left[u - (\ell +1) v_k\right] - G_0 \left[u + (\ell+1)  v_k\right]}{\omega - k  u} d u \\
& +i\pi \frac{m\omega_p^2}{\hbar k^2} \sum_\ell (\ell + 1)  \left\vert  J_\ell (\xi) \right \vert^2 \left[ G_0 \left(\frac{\omega}{k} - (\ell +1) v_k\right) - G_0 \left(\frac{\omega}{k} + (\ell+1)  v_k\right) \right].
\label{eq_landau}
\end{align}
Here, $\wp$ stands for the principal value of the integral, and the last line of Eq. \eqref{eq_landau} has been obtained with the help of the Plemelj formula, $\lim_{\epsilon \to 0}(x\pm i\epsilon)^{-1} =\wp x^{-1} \mp i \pi \delta(x)$. Finally, by expanding Eq. \eqref{eq_landau} at first order in $\gamma$, we obtain 
\begin{equation}
\gamma  = \frac{\pi \omega_p^3}{4 k^2} \sum_\ell \frac{\ell+1}{v_k} \left| J_\ell (\xi) \right|^2 \left[ G_0 \left( \frac{\omega}{k} + (\ell+1) v_k \right) - 
G_0 \left( \frac{\omega}{k} - (\ell +1) v_k \right) \right] \, .
\label{3.6b} 
\end{equation}
The features of the nonlinear damping rate are depicted in Fig. \ref{fig_imag} for a degenerate plasma, in which the electrons follow a Fermi-Dirac distribution,
\begin{equation}
G_0(\zeta)=\frac{1}{z^{-1}e^{\zeta}+1},
\end{equation}
where $\zeta=mu^2/(2T_e)$, $z=e^{E_F/T_e}$ is the fugacity and $E_F$ is the Fermi energy. As we can observe, for finite amplitude waves, Landau damping is suppressed, as consequence of electron trapping. Although somehow counter-intuitive $-$ one could naively believe that a finite amplitude wave tends to interact more effectively with single electrons and, eventually, damp its energy into the later $-$ this result is well aligned with previous results in the literature. In fact, the Bernstein-Greene-Kruskal (BGK) modes, a class of exact nonlinear solutions in collisionless plasmas, are known to be undamped \cite{Ng2006}. Moreover, our results provides analytical support to the previous numerical results of Brodin et al. \cite{Brodin2015, Brodin2017}, reporting on the relevance of multi-plasmon resonant wave-particle effects. \par
\section{Conclusions}

In this paper, we have developed a theoretical framework capturing the main aspects of the nonlinear Landau damping of electron plasma (Langmuir) waves in the quantum regime. By making use of a Volkov formulation of the problem, we are able to derive a formal solution to the test particle driven by a finite amplitude electrostatic wave of the plasma. The subsequent statistical, self-consistent treatment of the plasma electrons is obtained for Wigner function describing a mixture state, which evolves according to a quantum kinetic equation of the Wigner-Moyal type. The crucial difference with respect to standard applications of the Wigner-Moyal equation in the study of plasma waves stems in the possibility to consider finite amplitude waves, which now appears as a parameter in the dispersion relation. We show that, in the presence of finite amplitude waves, the nonlinear kinetic dispersion relation provides different modes, unlike to the case of Langmuir waves. Such modes are hybridizations between the (linear) Langmuir mode (now corrected by a quantity proportional to the amplitude of the wave) and the single-particle modes composed by the electrons that are not trapped by the wave. The latter corresponds to a mode quasi-particle which mass is controlled by the number of plasmons exchanged between the electrons. Moreover, we have shown that Landau damping in the nonlinear regime is strongly suppressed. This is, again, a consequence of particle trapping, which prevents particles to participate in the process of extracting energy from the wave. The calculations performed here are in agreement with the numerical results performed on the numerical simulation of Wigner-Moyal equation \cite{Brodin2017}, while keeping the simplicity of the formulation. This is a consequence of the versatility and power of the Volkov approach. \par
Given the versatility of the present approach, we believe that our findings will stimulate further discussions around nonlinear wave-particle interactions, in general, and around Landau damping, in particular, in a plethora of systems featuring long range interactions. First, there is a vast set of condensed-matter platforms in which this formalism can be applied and tested, namely two-dimensional plasmonics in graphene and related materials \cite{Gardner1994, Mustafa2010, Moradi2017, Figueiredo2022, Cosme2022, Cosme2023}, including bilayer graphene, black phosphorus and transition metal dichalgogenides (TMDCs) \cite{Khan2020}, one-dimensional plasmas in carbon nanotubes and metallic nanowires \cite{Chang2006, Zhang2011, Huang2023}, and three-dimensional Dirac materials such as Weyl semimetals \cite{Zhang2021}. Eventual differences in regard to this work come from the dispersive nature of the carries in the band theories, which must be taken into account in the derivation of both single-particle and kinetic (Wigner) equations. The strategy, however, remains the same. Second, in self-gravitating systems, numerous quantum gravity studies are performed with the help of Schr\"odinger-Poisson system, given the quantum mechanical nature of the fields composing dark matter. For instance, in Refs. \cite{Chavanis2011, Gomes2023} dark matter is treated as a Bose-Einstein condensate, while a quantum kinetic formulation of the problem has been put forward in Refs. \cite{Mendonca2019, Mendonca2021}. Moreover, classical self-gravitating phenomena, such as Jeans instability and the formation of structures, are usually described by a kinetic equation of the Vlasov type \cite{Farias2018, Moretti2020, Ourabah2020, Ourabah2020b}. Indeed, at a first glance, no quantum mechanics takes place here. The important point lies, however, in the fact that the Vlasov equation easily follows from the classical limit of the Wigner equation, which can be formally obtained by setting $\hbar \to 0$ in Eq. \eqref{3.2b}. Thus, multiple plasmon effects, such as those patent in Eq. (25) and subsequent equations are, in principle, possible in the context of self-gravitating phenomenology. More precisely, the Jeans length $\lambda_J$ establishes the typical length at which structures form; if nonlinear Landau damping occurs as we here predict for plasmas, then important changes may take in the correct definition of $\lambda_J$, which would now be determined through the competition between the Jeans instability (positive growth) and the damping (negative growth) of normal modes.

{\it Acknowledgments.}---F. H. would like to thank the Instituto Superior T\' ecnico and, in particular, the Group of Lasers and Plasmas (GoLP), for hospitality and support during his visit to Lisbon. He also acknowledges financial support of CNPq (Conselho Nacional de Desenvolvimento Cient\'ifico e Tecnol\'ogico), Brazil. H. T. acknowledges Funda\c{c}\~{a}o da Ci\^{e}ncia e a Tecnologia (FCT-Portugal) through Contract No. CEECIND/00401/2018 and Project No. PTDC/FIS-OUT/3882/2020. Stimulating discussions with Pedro Cosme are also acknowledged.

\bibliography{REFERENCES.bib}

\end{document}